\documentclass[journal]{IEEEtran}

\ifCLASSINFOpdf

\else

\fi

\usepackage{graphicx}

\usepackage{amssymb}
 \usepackage{amsthm}
\usepackage{amsmath}
\usepackage{color}
\usepackage{setspace}
\usepackage{algorithm}
\usepackage{algorithmic}
\usepackage{graphicx}
\usepackage{subfigure}
\usepackage{multirow}
\usepackage{array}
\usepackage[font=small,labelfont=bf,tableposition=top]{caption}
\usepackage{booktabs}
\usepackage{threeparttable}
\usepackage{stfloats}
\usepackage{color}
\usepackage{url}
\renewcommand{\baselinestretch}{1.0}

\begin{document}


\title{Food Recommendation: Framework, Existing Solutions and Challenges}

\author{Weiqing Min, \textit{Member, IEEE}, Shuqiang Jiang, \textit{Senior Member, IEEE}, Ramesh Jain, \textit{Fellow, IEEE}
\thanks{
This work was supported in part by the National Natural Science Foundation of China under Grant 61532018 and 61972378, in part by Beijing Natural Science Foundation under Grant L182054, in part by National Program for Special Support of Eminent Professionals and National Program for Support of Top-notch Young Professionals.
W. Min is with the Key Laboratory of Intelligent Information Processing, Institute of Computing Technology, Chinese Academy of Sciences, Beijing, 100190, China, and also with State key Laboratory of Robotics, Shenyang Institute of Automation, Chinese Academy of Sciences, Shenyang, 110016, China. email:minweiqing@ict.ac.cn.
S. Jiang is with the Key Laboratory of Intelligent Information Processing, Institute of Computing Technology, Chinese Academy of Sciences, Beijing, 100190, China, and also with University of Chinese Academy of Sciences, Beijing, 100049, China email: sqjiang@ict.ac.cn.
R. Jain is with Department of Computer Science, University of California, Irvine, CA, USA, email: jain@ics.uci.edu.
}
}
\markboth{IEEE Transactions on Multimedia,~Vol.~X,
No.~XX,~Month~Year}{}



\maketitle

\begin{abstract}
A growing proportion of the global population is
becoming overweight or obese, leading to various diseases (e.g.,
diabetes, ischemic heart disease and even cancer) due to
unhealthy eating patterns, such as increased intake of food with
high energy and high fat. Food recommendation is of
paramount importance to alleviate this problem. Unfortunately,
modern multimedia research has enhanced the performance and
experience of multimedia recommendation in many fields such
as movies and POI, yet largely lags in the food domain. This
article proposes a unified framework
for food recommendation, and identifies main issues affecting
food recommendation including incorporating various
context and domain knowledge, building the personal model, and analyzing unique food characteristics.  We then review existing solutions
for these issues, and finally elaborate research challenges and
future directions in this field. To our knowledge, this is the first
survey that targets the study of food recommendation in the
multimedia field and offers a collection of research studies
and technologies to benefit researchers in this field.
\end{abstract}


\IEEEpeerreviewmaketitle

\section{Introduction}

Food  is always central to  the human life. Besides the air we breathe, food is the only physical matter, which humans take into the body. In the early days, humans  faced the  task of identifying and gathering food for their survival. At present, the dietary choice is becoming vital in satisfying diverse needs, such as  basic  nutrition, calorie, taste, health and social occasions. According to the International Diabetes Federation, about 415 million people worldwide suffer from diabetes, and the rate of diabetes incidence is projected to further increase by more than 50\% by 2040, becoming one great threat to global health.   The dietary factor  is one main cause of  the dramatic increase in the incidence of obesity and diabetes~\cite{Tirosheaav0120,Federation-IDF2015}. The Global Burden of Disease Study also indicates dietary factors as a major contributor to levels of malnutrition, obesity and overweight, and unreasonable  diets lead to 11 million avoidable premature deaths per year~\cite{GBD-Global-Lancet2018}.  Food computing~\cite{Min2018A} is  emerging as a new field to ameliorate these issues. As an important task in food computing, food recommendation intends to find  suitable food items for users to meet their personalized needs,  and thus plays a critical role in  human dietary choice.

A balanced diet is crucial to maintain one's physical health. However, nutrients that need to be ingested vary greatly depending on  personal food preference and health conditions. Therefore, how to provide personalized food recommendation according to different personal requirements is very important. The past  decade has witnessed  the rapid growth of internet services and mobile devices. It has been more convenient for people to access huge amounts of online  multimedia food content from various sources, such as forums, social media, recipe-sharing websites and customer review sites. Although this growth allows users to have more choices, it also brings problems for users to  select  preferred food items from  thousands of candidates.  Therefore, food recommendation is becoming increasingly essential for serving  potentially huge service demand and can help users easily discover a small subset of food items which are enjoyable and suitable for them.

Compared with recommendation in other fields, food recommendation has its own characteristics. For example, food preference learning is an important step towards food recommendation. However, food preference involves  various factors, such as taste preference, perceptive difference, cognitive restraint, cultural familiarity and even genetic  influence\footnote{\url{https://www.eurekalert.org/pub_releases/2017-04/eb2-cgi041217.php}}. Therefore, it makes  accurate food preference learning more difficult. Furthermore, food recommendation should consider more context information. Besides basic context information captured from familiar mobile devices~\cite{Weiqing-CAMVR-MSJ2017}, such as time, location and environmental information (e.g., temperature and PM2.5), various body state-related signals, such as steps taken, heart rate, sleep quality, body acceleration and even affective states can also be captured from  new sensing devices, such as watches, wearable fitness trackers and bracelets~\cite{Boll-HM-IEEEMM2018}. These signals can describe users' actual body conditions comprehensively in real-time  and is of crucial importance for food recommendation. However, they are with different statistical properties, either discrete or continue, and  make effective multi-sensor context fusion  challenging. As a result, simply borrowing methods from recommendation methods in other fields without considering characteristics of food recommendation probably leads to inferior performance. This negligence will  eventually result in a bottleneck in advancing intelligent food recommendation for use in real-world applications.

Existing multimedia research has made great progress in improving the recommendation performance and experience in many fields such as movies and POI, yet largely lags in the food domain.  To the best of our knowledge, although relevant works on food  recommendation \cite{Maruyama-RMRR-MM2012,Nag2017Live,Nag-HML-ICMR2017} have received more attention in the multimedia community, there are very few systematic reviews, which provide a unified framework and comprehensive summary of current efforts in food recommendation. Because of huge potentials in food recommendation, the time has come for the multimedia field to give a survey on food recommendation, which can help  researchers from  relevant communities better understand the strength and weakness of existing methods. In this paper, we give a survey on food recommendation. Particularly, the objective of this article is as follows: to propose a unified framework  for food recommendation  and identify  main issues affecting food recommendation (Section~\ref{framework}); to review  existing progress  for these issues (Section~\ref{Existingsolutions}), and  to outline research challenges and future directions in this field (Section~\ref{Challenge} and Section~\ref{Directions}). Finally, we conclude the survey in Section ~\ref{Conclusions}.

Note that recently there is one  survey work on food recommendation~\cite{Trattner2017Food}. Our survey and \cite{Trattner2017Food} both summarize existing methods, challenges and future directions, and  are both very complementary to benefit researchers and practitioners working in this field. However, there are  three important differences between this work and \cite{Trattner2017Food}: (1) To our knowledge, we first systematically define the problem of food recommendation and summarize its three  unique aspects. (2) We propose a unified food recommendation framework, and  put previous studies together under this unified umbrella. Consequently, our taxonomy on existing food recommendation works  is different from \cite{Trattner2017Food}. (3) We put more emphasis on multimedia-oriented food recommendation, where how to utilize and fuse multi-modal signals (e.g., images,voice and text) and rich context for food recommendation is highlighted. In contrast, \cite{Trattner2017Food} mainly focuses on food recommendation based on recommendation techniques.

\section{Proposed Framework}\label{framework}

\begin{figure*}
\centering
\includegraphics[width=0.95\textwidth]{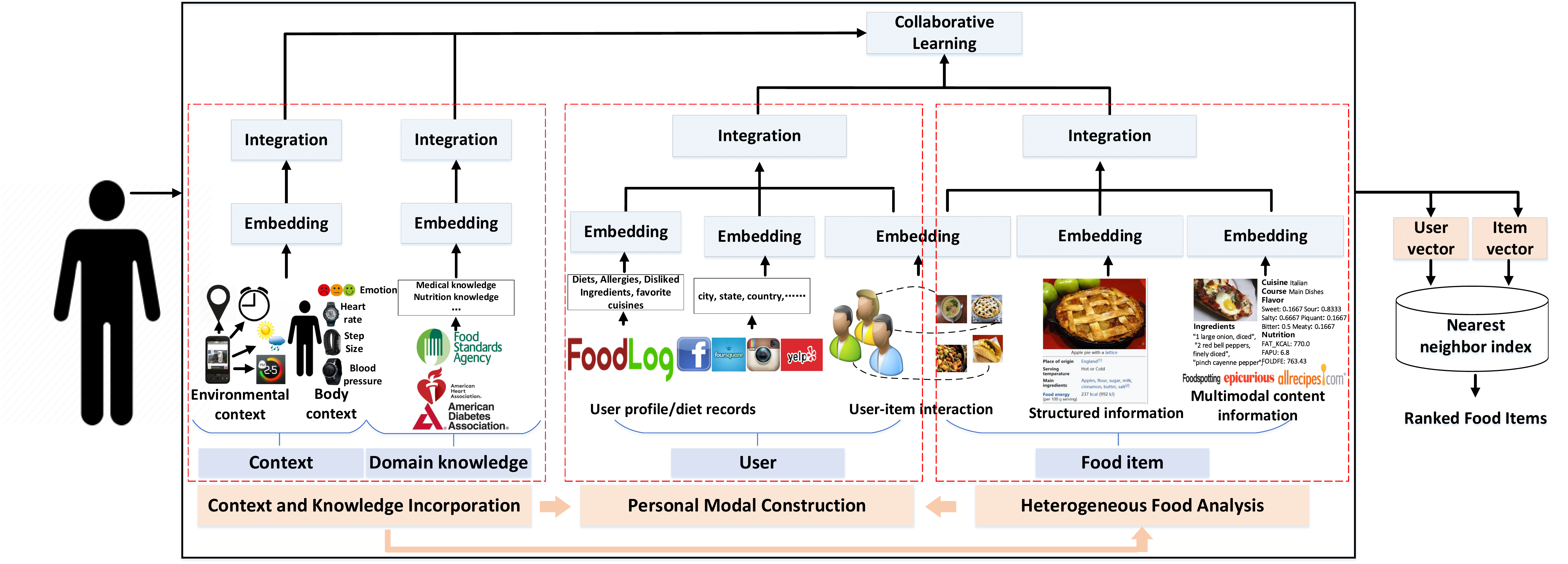}
\caption{The general framework of food recommendation, where collaborative  learning is adopted to learn both user latent vectors and item latent vectors after embedding and integration on context-dimension, user-dimension and food item-dimension, respectively. For each dimension, the signals from different sources are first changed into the feature representation via embedding, and  are then fused into final representation via integration.}
\label{food recommendation_solution}
\end{figure*}

\textbf{Definition} Food recommendation aims to provide a list of ranked food items for users to meet their personalized needs. Here, food is a more broad concept, and it includes all food-related items, such as  meal, recipes, coffee shops and restaurants. Food recommendation is typically multidisciplinary research, including nutrition, food science, psychology, biology, anthropology, sociology, other branches of  natural and social sciences\footnote{\url{https://www.wur.nl/en/show/Food-preferences.htm}}.

Compared with other types of recommendation,  there are mainly three aspects unique for food recommendation. (1) Food recommendation involves different context and domain knowledge. Rich user context (e.g., the heart rate and steps taken) and external environmental context (e.g., physical activity-relevant and health-relevant context) captured from different sensors describe users' actual physical conditions and their surroundings, and thus provide valuable information for exact match between user requirement and food items. For example, food recommendation probably recommends one user food items with much water and protein after exercise captured from sensors. Furthermore, food recommendation is very relevant to health. Therefore, medical knowledge, dietary knowledge and other relevant domain knowledge should also be incorporated into the food recommender system for constraint optimization and computing. (2) From the user aspect, the most significant difference is that food recommendation is very relevant to user's health. Therefore, an ideal food recommendation system should self-adaptively build a trade-off between personalized food preference/interest and personalized nutrition/health requirement. For example, even one diabetic  likes sweet food, it is more reasonable for food recommendation to recommend him/her the food with less sugar than before. Besides user's health needs, food recommendation should also consider other complex and various fine-grained user needs, such as allergies and life-style preferences (e.g., the desire to eat only vegan or vegetarian food). Therefore, a personal model should be built to take these factors into consideration. (3) Food has its own characteristics, and  many  unique factors, such as cooking methods, ingredient combination effects, preparation time, nutritional breakdown, and  non-rigid visual appearance should be analyzed to obtain high-level semantic concepts and attributes for  food recommendation. For example, nutritional and calorie  intake  can be calculated by analyzing captured food images to create a quantitative nutrition diary. Therefore, multimodal heterogeneous food analysis is necessary to accurately obtain  high-level understanding of the food type and other levels of semantics.

Taking all these factors into consideration,  we propose a unified framework (as shown in Fig.~\ref{food recommendation_solution}), which can jointly utilize  rich  context and knowledge, user  information, and heterogeneous  food information for food recommendation. In order to combine these content information with user-food item interaction, a hybrid recommendation strategy is utilized \cite{Burke2002Hybrid} in the food recommendation framework, where joint collaborative  learning is adopted to learn both user latent vectors and item latent vectors after embedding and integration on context-dimension, user-dimension and food item-dimension, respectively. For each dimension, the signals from different sources are first changed into the feature representation after embedding, and  are then fused into final representation via integration, where different embedding  and integration methods can be adopted. Final recommendation is generated from these learned user vectors and item vectors under context constraints. Particularly, food recommendation mainly consists of three components, namely context and knowledge incorporation, personal modal construction and heterogeneous food analysis. Note that the context and knowledge can not only put more constraints on food recommendation independently, but also be used to support the personal model construction and heterogenous food analysis. In addition, heterogenous food analysis is also helpful for building the complete personal model.  Next, we will identify detailed requirements of three components, respectively.
\begin{figure}
\centering
\includegraphics[width=0.45\textwidth]{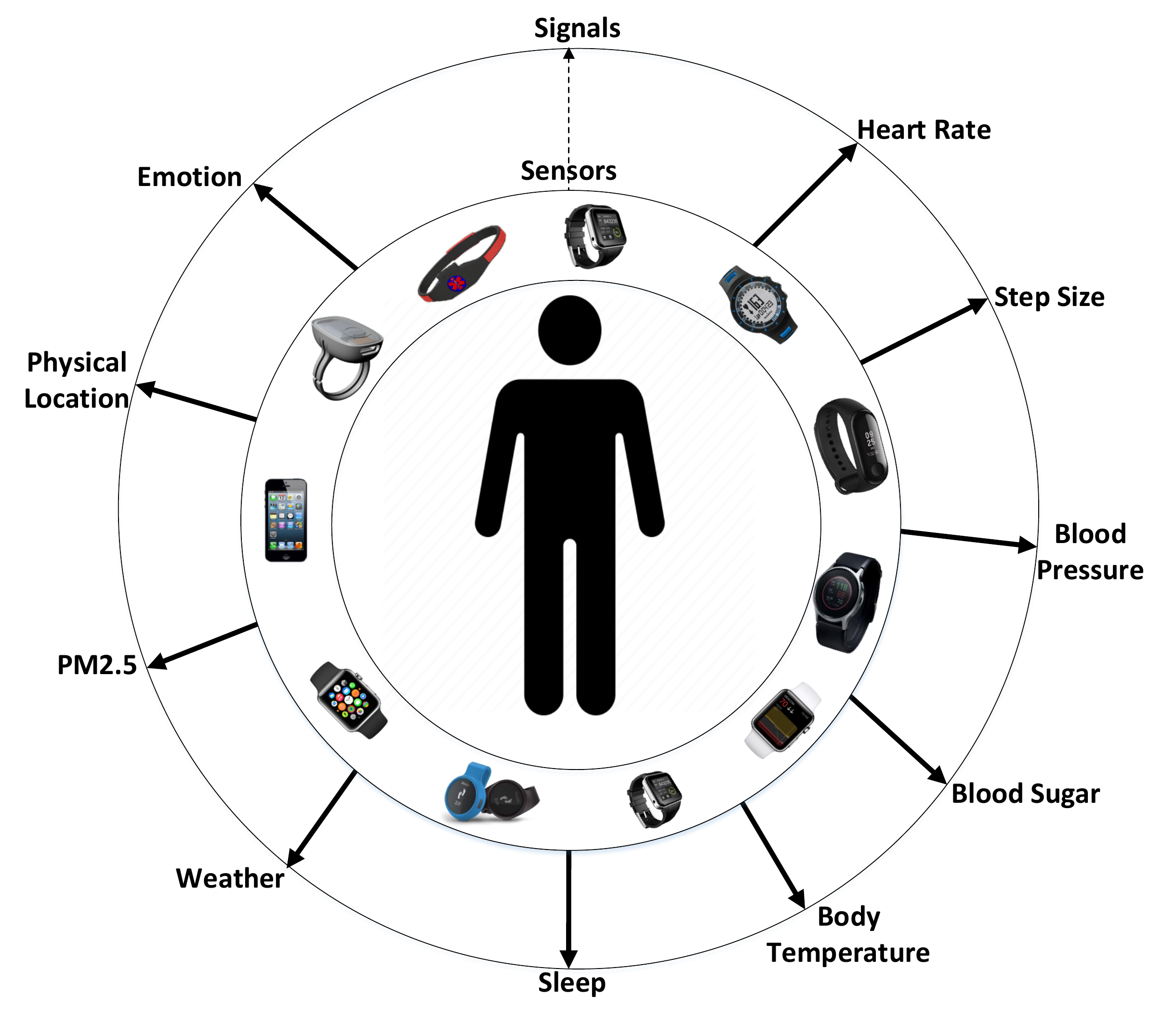}
\caption{Various signals from  sensor devices.The new sensing devices can measure physiological parameters to produce various signals, such as step sizes, heart rate, sleep quality, body acceleration, blood pressure and even emotion.}
\label{food recommendation_body_state}
\end{figure}
\subsection{Context and  Knowledge Incorporation}
Basic context information (e.g., time and location) can help filter out irrelevant  items for recommendation. Compared with other types of recommendation, food recommendation  involves more complex, diverse and even dynamic factors. Rich user context and external environmental context information describe users' actual physical conditions and their surroundings, and  thus provide valuable information for exact match between user requirement and food items of interest.

Over the last decade, a great variety of wearable electronic devices and ambient sensors have been developed \cite{Lou2017Recent}. They can monitor personal body conditions and environmental changes everywhere in real time by connecting  users  to surrounding machines. The new sensing devices, ranging  from social sensors to biosensors \cite{Boll-HM-IEEEMM2018} can measure physiological parameters to produce various  signals, such as step sizes, heart rate, sleep quality, body acceleration, blood pressure and affective states (Fig.~\ref{food recommendation_body_state}). In addition, various  health and fitness mobile apps such as MyFitnessPal, Endomondo and Fitbit can also help people keep track of what they eat, when they exercise, and how well they sleep. Exploiting these context information will enable more reasonable and accurate food recommendation. For example, if the level of sugar captured from the sensor is high for one user, it is necessary  to recommend this user the food with little sugar or low conversion sugar. Food recommender system probably recommends one user the food with much water and protein after exercise, captured from sensors. Besides rich context, health-relevant expert knowledge (e.g., medical knowledge and nutrition knowledge) should also be considered. Food recommendation via expert knowledge is a potential key to unlock healthy diets, and thus gives more precise recommendation results.

There are mainly two  issues to solve when applying  rich context and knowledge for food recommendation: (1) Current wearable sensing technologies  face many challenges, such as inaccuracy and uncertainty of  measurement, multi-functional integration and adverse impacts on the environment. Therefore, there are still many body-related signals, which are harder to capture accurately or even not available. More new wearable sensing technologies should be developed  to  obtain comprehensive human signals relevant to food recommendation. For example, researchers recently designed new sensors to detect  muscle motions involved in every chew, and finally can accurately create a time-stamped visual record of the food consumed~\cite{Strickland-Sensors-IEEESp2018}. (2) Multiple-sensor context  predominantly is segregated with the taxonomy of categorical/discrete and ordinal/continuous features in time and space. For example, some signals such as step count is discrete while others  such as electrocardiogram are continuous. In addition, the categorical  features probably vary widely in their cardinality.  Some are binary while others have many possible values. It is difficult to  model user states  to accommodate a mixture of discrete and continuous variables in a joint model, not to mention  external knowledge. Inappropriate  representation and fusion methods for rich context and knowledge are even not as effective as  methods without any context. In addition, the variety and number of contexts makes it hard to measure all available data, which in turn introduces new uncertainty levels for food recommendation. Therefore, an effective multi-sensor fusion method should be developed to overcome these problems.

\subsection{Personal Model Construction}

Personal model construction involves collecting and fusing relevant information of users to generate a user  model for  food recommendation.  A food recommendation agent cannot function accurately until the personal model has been well constructed. The system needs to know the information from the user as much as possible in order to provide reasonable and precise recommendation.

A user profile is a collection of personal information associated with a specific user, and  can be considered as a simple personal model. In contrast, food recommendation  considers more complex, dynamic and various constrained factors, such as attitudes and beliefs about food and recipes, personal food preference, lifestyles, hobbies, and even cultural and social aspects. In order to effectively construct the personal model for food recommendation, we should capture and fuse various signals from above-mentioned sensors, websites and social media. People have started recording health-related information using sensors ranging from simple wearable accelerometers that could be classified and recorded in simple activities (such as walking, jogging and climbing stairs) to other measurements (such as body temperature, heart rate, perspiration rate, galvanic skin resistivity, and many other deeper parameters). We can use this to model  user's real-time states continuously. In addition, we are leaving digital traces of all kinds of activities online in the  social media. As food is central in our life, a significant fraction of online content is about food. A huge amount of information about eating habits is now being recorded digitally and available.  Besides shared multimedia food content, we can obtain both basic user information (e.g., the age, gender and residence) and more detailed information (e.g., diets and allergies). For example, we can obtain fine-grained information from Yummly, such as diets (e.g., vegetarian or low fodmap), allergies (e.g., seafood or tree nut), disliked ingredients (e.g.,sugar or beef), favorite cuisines (e.g., Chinese and Indian cuisine), taste (e.g., sour, salty, sweet, bitter and meaty) and food restrictions (e.g., vegetarian, vegan, kosher and halal). All these user information can also be taken into consideration for personal model construction via heterogeneous media processing.

Among these different data sources, Foodlog is an important source for personal model construction. Foodlog has recorded  users' food intake by taking photos of their meals or writing the diary  for food  assessment and journaling. Traditional methods of keeping a food journal resort to manually recording meals in as much detail as possible by including the portion size, number of servings and calories, time, location, or even the people around us. Currently, multimodal food journal is developed \cite{Aizawa-FoodLog-IEEEMM2015}. The rich and detailed recording information provides us with fine-grained and accurate user preference on food, and finally leading to accurate personal model construction. However, to the best of our knowledge, few works  exploits rich structured information from Foodlog to learn personalized food preference for  personal model construction.

Besides complex food preference, personal nutritional and health factors should also be emphasized in building the personal model. For example, nutritional fitness offers an objective way to prioritize recommended foods for each physical and dietary condition \cite{Kim2018Nutritionally}. That is, food recommendation should   consider not only  what he likes, but also what is nutritionally appropriate. When we incorporate  nutrition information into food recommendation, it should be  also personalized. Each user's nutrition is complex and uncertain on many levels \cite{Elahi2017User}. This contrasts with most public nutrition and health advice, which is generic, non-specific healthy eating advice e.g., eat at least five portions of fruit and vegetables daily. Therefore, how to accurately build a personal model  is an important issue, and should be deeply explored.

\subsection{Heterogeneous Food Analysis}
Food analysis is one basic component of food recommendation, as it provides  prerequisites for obtaining a high-level understanding of the type (e.g., food category and ingredients), the volume of consumed food,  nutrition and calorie intake by the user. Particularly, heterogeneous food analysis can be considered from the following three aspects.

(1) Effective representation of single modality. Take food images as an example, with the prosperity of mobile services, more and more users tend to take photos they eat instead of  text recording. Food images  has become more prevalent. Visual analysis  provides not only  ingredient and calorie information, but also  visual perceptual information. It has proved that visual information is helpful for  food preference learning \cite{Yang2017Yum}. However, it is very difficult for fine-grained visual food representation because of their  indistinctive spatial configuration and various types of geometrical variants~\cite{Bossard-Food101-ECCV2014}. Different from general object recognition, many types of food do not exhibit distinctive spatial  configuration. They are typically non-rigid, and the structure information can not be easily exploited. Therefore, existing recognition approaches on general objects probably do not perform well for lack of  effective visual food representation.

(2) The  multimodality of food data. The food  actually involves multi-modalities. We see food, feel its texture, smell its odors, taste its flavors and even hear its sounds when chewing. For example, when we bite into an apple, not only do we taste it, but also hear it crunch, see its red skin, and feel the coolness of its core~\cite{Smith2014The}. Visual information of a food product such as  color and texture can exert an influence on the acceptance of  food products \cite{Cordeiro-RMFJ-HFCS2015}. Furthermore, with the explosive growth of multimedia food content and service, food items  consist of different  modalities or structures, such as food images, cooking videos, ingredient lists, cooking instructions and various attributes. We should fuse different types of modalities for effective food representation. Despite these advances, multimodal food fusion still faces some challenges. For example, signals might not be  aligned, such as continuous food image signals and discrete ingredient signals. Each modality might exhibit different types and different levels of noise at different points.

(3) The  heterogeneity of food data. Food data has also different structures, and thus contains richer semantic information. For example, in food-oriented Knowledge Graph (KG), entities usually include food items and relevant attributes (e.g., ingredients and taste), and the relationship between entities (e.g., belong to ``Kung Pow Chicken belongs to Chinese cuisine", and include  ``Chow mein includes noodles, soy sauce and vegetables ingredients"). KG can benefit the recommendation in many aspects, such as  improving the precision and diversity of recommended items. Because the dimension of knowledge graph is higher and the semantic relationship is richer, the processing of KG is more complex and difficult.

\section{Existing solutions}\label{Existingsolutions}
Taking the above-mentioned  issues into consideration, this section first surveys existing solutions from the following three aspects, corresponding to each component of the proposed food recommendation framework: 1) incorporating  context and knowledge for food recommendation, 2) personal model construction for food recommendation, 3) heterogeneous food analysis for food recommendation. In addition, considering the importance of nutrition and health in food recommendation, we therefore surveys existing solutions  on 4) nutrition and health oriented food recommendation.

\subsection{Incorporating  Context and Knowledge for Food Recommendation}
\begin{figure}
\centering
\includegraphics[width=0.40\textwidth]{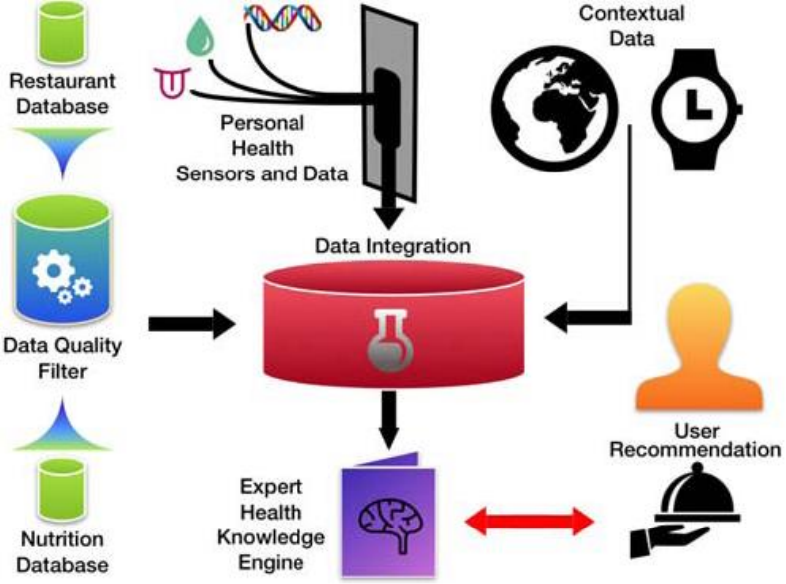}
\caption{Live context aware personalized dietitian system~\cite{Nag2017Live}.}
\label{foodcontext}
\end{figure}

Food recommendation is highly dependent on current context. Any such context would need to be detected either by direct user input or sensing  environmental variables via various sensors, such as  GPS and accelerometers to heart-rate monitoring and blood oxygen level. The location context has been widely used for food recommendation \cite{Yanjie-UPL-SIAM2014,NagPSLWJ17,Nag2017Live}. For example, Yanjie \textit{et al.}~\cite{Yanjie-UPL-SIAM2014} integrated the geographic proximity into one generative probabilistic model to capture the geographic influence for restaurant recommendation.  Another important dimension is time, and it is very relevant to food recommendation because of  food popularity~\cite{Sanjo2017Recipe}. Nag \textit{et al.} \cite{Nag2017Live} further fused more types of context data via one pre-defined equation, such as  barometer and pedometer output to  estimate user's daily nutritional requirements for local dish recommendation.  The emerging sensor technologies such as social sensors and biosensors ~\cite{Meyer-DHDE-DPC2014,Boll-HM-IEEEMM2018}  provided richer sensing context, especially health-relevant one. These context information put more constraints on recommendation to  improve the accuracy of recommendation system.

Besides rich context from various sensors, efforts in modeling expert knowledge are made. For example, nutrition facts can be readily available for all major restaurant chains. For packaged items, algorithms that use this information are most promising for immediate consumer use and health impact. USDA Food Composition Databases\footnote{\url{https://ndb.nal.usda.gov/ndb/search/list}}, North American derived Nutrient Rich Foods Index 6.3 (NRF)~\cite{Fulgoni-MNQF-JN2009} and British FSA~\cite{Julia-VFSA-EJN2016}  have been built based more heavily on available nutrition facts. However, they  have not been established to capture expert knowledge of dietitians.

As representative work, Nag \textit{et al.} \cite{Nag2017Live} proposed a  live personalized nutrition recommendation system. As shown in Fig. \ref{foodcontext}, a decision support system is made by fusing  timely, contextually aware, personalized data to find local restaurant dishes to satisfy  user's needs. Multi-modal contextual data including GPS location, barometer, and pedometer output is fused through one pre-defined equation for calculating daily values. The calculated daily values are then used to rank the meals for local meal recommendation.

Discussion. In this section, we identify different types of context and their combination for food recommendation. The common context such as GPS  and time information has been extensively used for food recommendation. One simple and effective method  is to directly use these context to filter out irrelevant  food items as the constraints, such as \cite{NagPSLWJ17,Nag2017Live}. Besides common spatio-temporal context, we can obtain richer  context information, such as the physical state and health state captured from various sensors. In this case, a joint model is necessary to  accommodate a mixture of both discrete and continuous context variables, such as~\cite{Yanjie-UPL-SIAM2014}. However, the noise generated from these sensors  makes effective context modeling more difficult. As for the domain knowledge, one reasonable method is to construct the food knowledge graph for food recommendation. However, there are few food knowledge graphs available. As a result, food knowledge has not been fully exploited by existing methods.

\subsection{Personal Model Construction for Food Recommendation}
\begin{figure}
\centering
\includegraphics[width=0.50\textwidth]{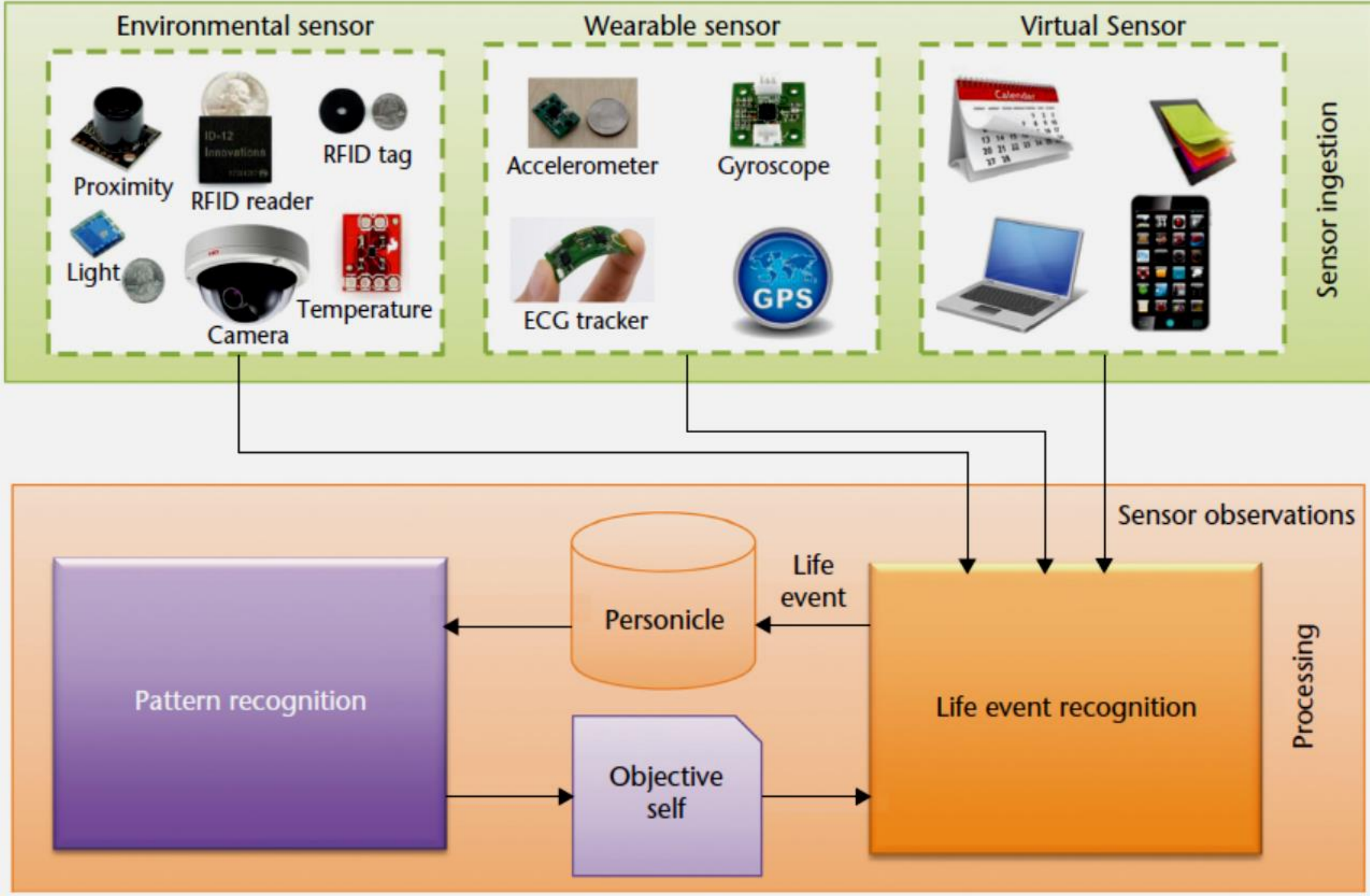}
\caption{High-level architecture of personal model construction system~\cite{Jain-IEEEMM2014}.}
\label{quantifiedself}
\end{figure}

The important step in making effective food recommendation is to build a personal model. Few works focus on personal model construction. The earlier work \cite{Jain-IEEEMM2014} proposed a high-level architecture of an objective self system for quantified self, namely personal model construction. As shown in Fig. \ref{quantifiedself}, the proposed architecture consists of  three main components data ingestion, life event recognition and pattern recognition. Data ingestion uses various sensors and preprocessing modules to extract appropriate attributes. Life event recognition predicted the most appropriate life activity from a set of predefined event classes based on pattern recognition. Later, Nag \textit{et al.} \cite{Nag2017Live} further  utilized the objective self \cite{Jain-IEEEMM2014} for lifestyle recommendation.

In building the personal model, learning user's food preference is vital  as it provides personalized information for effective food targeting and suggestions. Users' food preference learning is mainly conducted via various types of interactions, including survey based methods,  web activity (e.g., ratings, browsing history and implicit feedbacks) based methods, online learning methods, food log based methods and dialogue based methods.

A typical on-boarding survey method should ask a number of multi-choice questions about general food preferences. Such methods are generally adopted by  many commercial systems, such as Zipongo\footnote{\url{https://meetzipongo.com/}}, Shopwell\footnote{\url{http://www.shopwell.com}} and PlateJoy\footnote{\url{https://www.platejoy.com/}}. For example, a daily meal planner app PlateJoy elicits preferences for healthy goals and dietary restrictions with some questions, such as ``Are there any ingredients you prefer to avoid? avocado, eggplant, eggs, seafood......". Yummly also asks users to select  allergies (e.g., seafood or tree nut), disliked ingredients (e.g., sugar or beef) and favorite cuisines (e.g., Chinese and Indian cuisine). Survey based preference elicitation methods are generally coarse-grained, and it cannot comprehensively capture user's preference.

Current popular methods resort to hybrids of historical records and  item ratings for food preference learning \cite{Freyne-IFP-ICIUI2010,Ge2015Using}. For example, Mouzhi \textit{et al.} \cite{Ge2015Using} extended matrix factorization by including additional parameters used for modelling the dependencies between assigned tags and ratings. For matrix factorization, let $\mathbf{p}_{u}\in R^{k}$ and $\mathbf{q}_{m}\in R^{k}$ denote  the vector of user $u$ and recipe $m$. In matrix factorization, ratings are estimated by computing the dot product of  vectors $\hat{r}_{um}=\mathbf{p}^{T}_{u}\mathbf{q}_{m}$. After introducing the tags assigned to the user and recipe into the model, ratings are now estimated as follows:
\begin{equation}\label{eq_MF}
\begin{aligned}
\hat{r}_{um}=(\mathbf{p}_{u}+\frac{1}{|T_{u}|}\sum_{t\in T_{u}}\mathbf{x}_{t})^{T}(\mathbf{q}_{m}+\frac{1}{|T_{m}|}\sum_{s\in T_{m}}\mathbf{y}_{s})
\end{aligned}
\end{equation}
where $\mathbf{x}_{t}$, $\mathbf{y}_{s}$ denote the feature vector from  user's and recipe's tags, respectively. $T_{u}$ and $T_{m}$ are the set of tags assigned by the user $u$ and recipe $m$, respectively.

\begin{figure}
\centering
\includegraphics[width=0.47\textwidth]{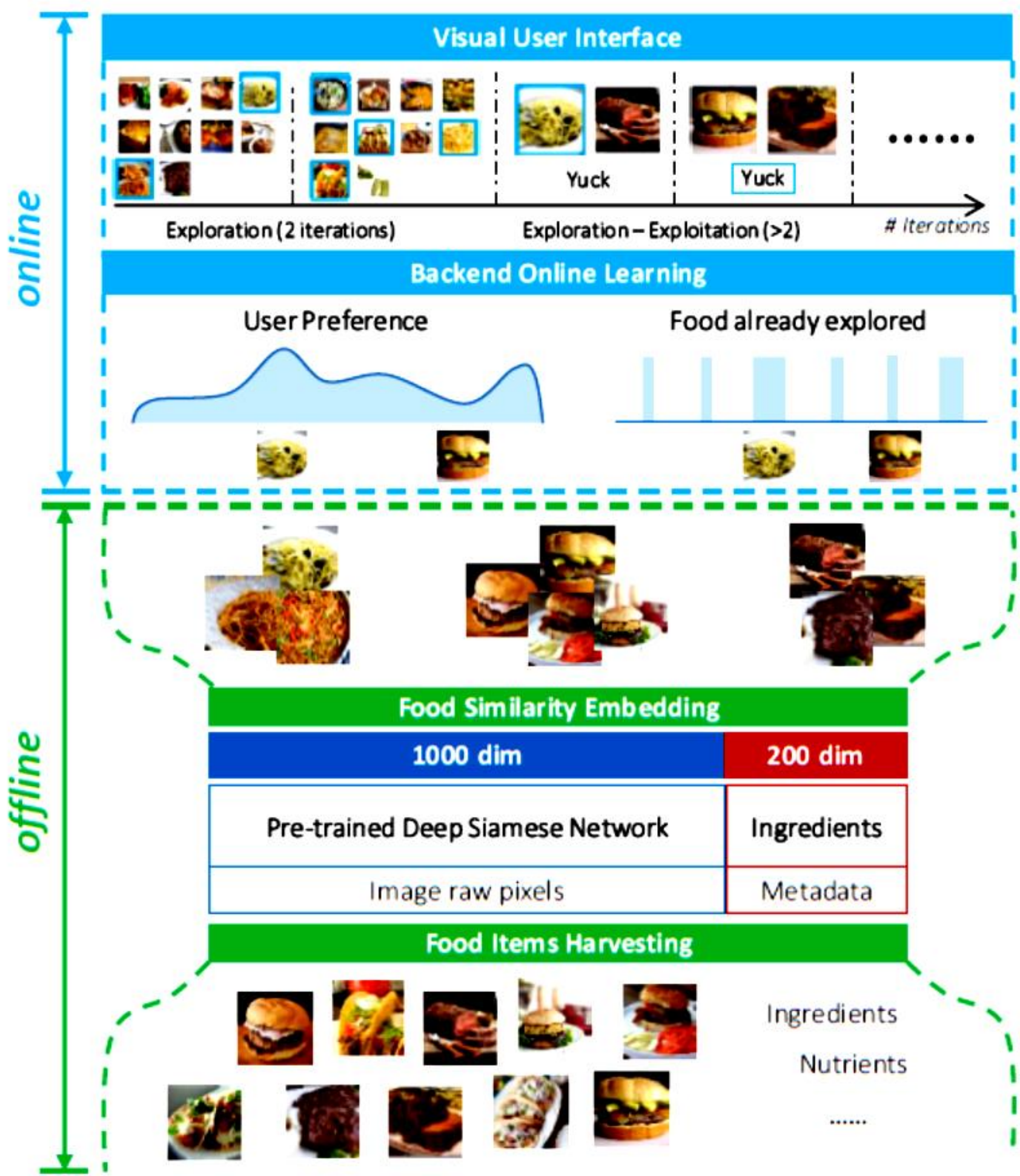}
\caption{PlateClick system pipeline~\cite{Yang-PlateClick-MM2015}.}
\label{food PlateClick}
\end{figure}

However, such methods suffer from the scarcity of user feedback. In contrast, Some works \cite{Yang-PlateClick-MM2015,Yang2017Yum,Gao-HANVFR-TMM2019} resorts to using visual content to learn user's food preference. For example, Longqi  \textit{et al.} \cite{Yang-PlateClick-MM2015,Yang2017Yum} learned users' fine-grained food preference with only visual content. They proposed a  food recommendation system PlateClick (Fig.~\ref{food PlateClick}), which consists of two stages, namely offline visual food similarity embedding and online food preference learning. In the first stage, a deep Siamese network is trained for visual food similarity embedding from  pairwise food image comparisons. This network is to learn a low dimensional feature embedding that pulls similar food items together and pushes dissimilar food items far away.  Contrastive loss is selected as the loss function  and can be expressed as:
\begin{equation}\label{eq_RA}
\begin{aligned}
&L=\{\underset{({i},{j})}{\sum}[yL_{o}(\mathbf{x}_{i},\mathbf{x}_{j}) +(1-y)L_{s}(\mathbf{x}_{i},\mathbf{x}_{j})]\}\\
\end{aligned}
\end{equation}
where $L_{o}(\mathbf{x}_{i},\mathbf{x}_{j})$ and $L_{s}(\mathbf{x}_{i},\mathbf{x}_{j})$ denote the contrastive constraint for ordered image pairs and the similar constraint for un-ordered image pairs, respectively. When incorporating  deep feature learning into Eqn. \ref{eq_RA}, they are denoted as:
\begin{equation}\label{eq_ManifoldR}
\begin{aligned}
&L_{o}(\mathbf{x}_{i},\mathbf{x}_{j})=\frac{1}{2}(\max(0,1-(f(\mathbf{x}_{i})-f(\mathbf{x}_{j})))^{2}\\
&L_{s}(\mathbf{x}_{i},\mathbf{x}_{j})=\frac{1}{2}(f(\mathbf{x}_{i})-f(\mathbf{x}_{j}))^{2}\\
\end{aligned}
\end{equation}
$f(\mathbf{x}_{i})$ is one low-dimensional feature embedding for $\mathbf{x}_{i}$. $y\in \{0, 1\}$ indicates whether the input pair of food items $\mathbf{x}_{i}$, $\mathbf{x}_{j}$ are similar or not ($y = 0$ for similar, $y = 1$ for dissimilar).

In the second stage, a novel online learning framework  is  explored  for learning users' preferences  based on a small number of user-food image interactions. The core of online learning lies in the preference propagation in locally connected graphs. Because of the subjectivity and uncertainty of visual perception from users, such methods can not accurately learn  food preference.

Recently, there are also some new methods for food preference learning. For example, Jie Zeng \textit{et al.} \cite{Zeng-Zeng-MHFI18} proposed a dialogue system to elicit user's food preference via human-robot interaction. To guarantee the conversation about food, the taste/texture expressions about dishes/ingredients are extracted from Twitter, and are  registered as the default knowledge base of the dialogue system.

Food journal/diary/log  records  users' food intake and daily diet by taking photos of their meals or writing the diary in detail by including the portion size, number of servings and calories, time, location, or even the people around us~\cite{Cordeiro-RMFJ-HFCS2015}.  This detailed description is effective, and thus we can learn users' food preference based on these historical records. For example, Nutrino\footnote{\url{http://nutrino.co/}}, a personal meal recommender, asks users to log their daily food consumption and learn users' fine-grained food preferences. As is typical of systems relying on user-generated data, food journaling suffers from the cold-start problem, where recommendations cannot be made or are subject to low accuracy when the user has not yet generated a sufficient amount of data.  This is because  traditional methods of keeping a food journal are manually recorded and it is very easy to forget or procrastinate logging food entries. In order to solve this problem, some methods have been proposed via  image recognition to recognize food items. For example, Foodlog~\cite{Aizawa-FoodLog-IEEEMM2015} has contributed to a record of users' food intake simply by taking photos of their meals. Recently, multimodal Foodlog~\cite{Hyungik-MFJ-ACMMM2018} is proposed to automatically recognize the starting moment of eating, and then prompt the user to begin a voice command food journaling method. With the popularity of mobile devices and advanced sensing technologies, learning user's food preference via exploring food logs  will be one effective way to accurately learn fine-grained user's food preference. Foodlog contains richer and more comprehensive dietary records, exploring Foodlog for accurate personal model construction will be a promising direction. However, because of  heterogeneous multimodal  content from Foodlog, it is not easy to build the personal model based on the Foodlog.

Discussion. Accurate personal model construction for food recommendation  is still challenging because of its  complexity of modeling and the diversity of factors to consider. As one important factor for one personal model, food preference learning has been widely exploited for food recommendation via various ways, such as survey based methods, historical records based methods  and dialog-based methods. Some  technologies such as matrix factorization and  deep Siamese network are utilized. However, the former suffers from  the scarcity of user data while the latter needs to construct a lot of item pairs. On the other hand, Foodlog comprehensively  records users' food intake, daily diet and other user profiles. Therefore, Foodlog-oriented food preference learning is expected to be very promising in the future.

\subsection{Heterogeneous Food Analysis for Food Recommendation}

\begin{figure}
\centering
\includegraphics[width=0.48\textwidth]{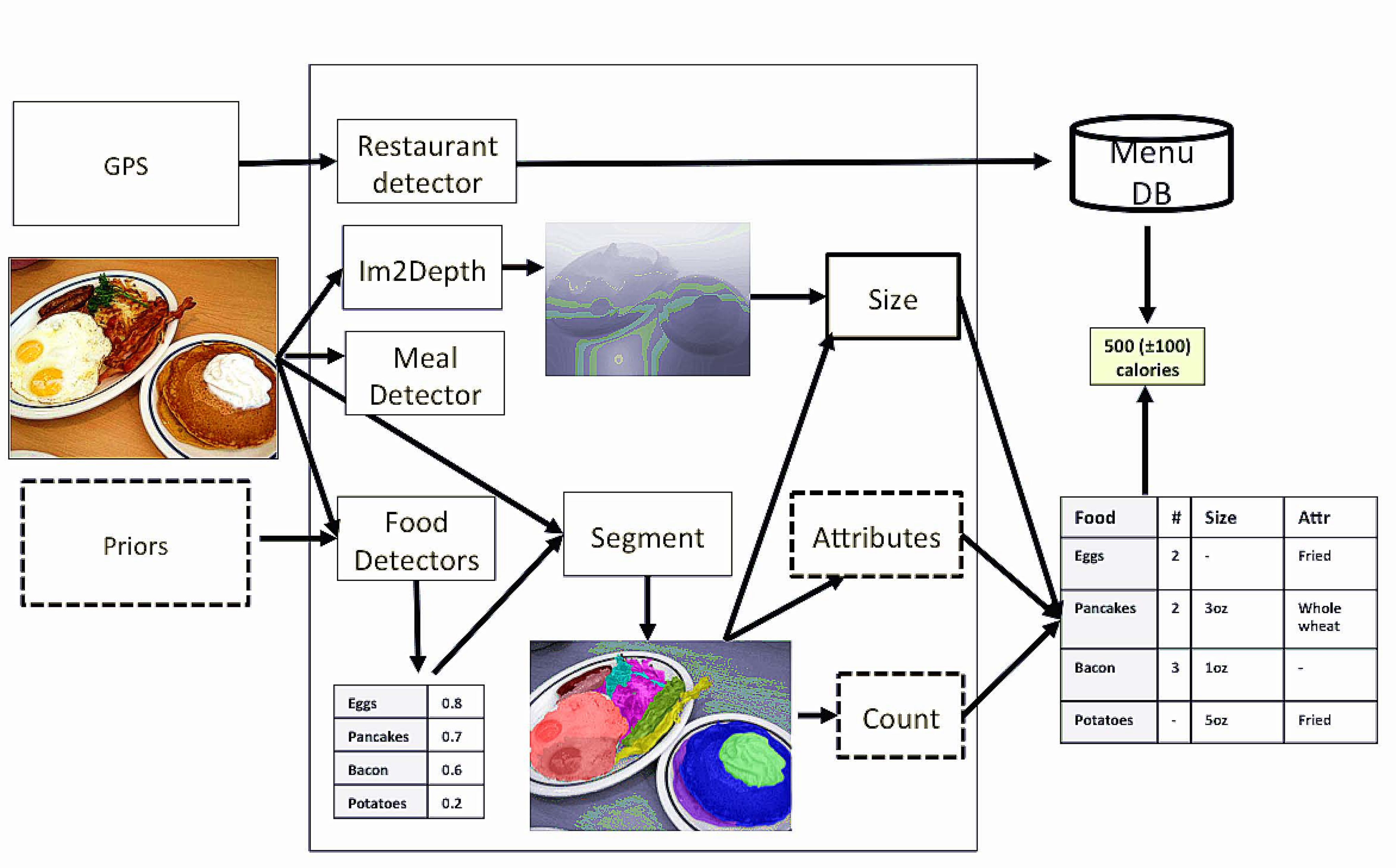}
\caption{Illustration of the overall system~\cite{Meyers-Im2Calories-ICCV2015}.}
\label{im2Calorie}
\end{figure}

For single modality-oriented representation, the basic analysis method is visual food recognition with or without context information \cite{Maruyama-RMRR-MM2012,Wang2016Where,Horiguchi-PCFIR-TMM2018,Aizawa-FBE-TMM2013,XuRuihan-GMDR-TMM2015,Chen-DIRCRR-MM2016}. We can obtain a high-level understanding of the type and the amount of food consumed by the user via food recognition. It has been used in Foodlog \cite{Aizawa-FBE-TMM2013}, food preference elicitation~\cite{Harvey2017Exploiting, Yang2017Yum} and  dietary tracking \cite{Ming-FPR-MMM2018} for further food recommendation. As  representative work, Myers \textit{et al.}~\cite{Meyers-Im2Calories-ICCV2015} presented a system which can recognize the content of the meal from one image, and then predicted its nutritional content, such as calories. It used a segmentation based approach to localize the meal region of the food photo, and then applied CNN-based multilabel classifiers to label these segmented regions. Once the system segmented the foods, it can estimate their volume. There are also other works for food volume estimation via the 3D model based on multiple food images \cite{Dehais2017Two}.

Besides  visual information, there are other types of content information available, such as ingredients and various food attributes for food recommendation. For example, Peter \textit{et al.}~\cite{Forbes-CBMF-RS2011} incorporated ingredient information as linear constraints to  extend the matrix factorization model for recipe recommendation. Devis \textit{et al.}~\cite{Devis-SDFR-WISE2015} selected relevant recipes via the comparisons among features used to annotate both users' profiles and recipes. Min \textit{et al.}~\cite{Min-YAWYE-TMM2018} considered one recipe as one document and one ingredient  as one word, and proposed a probabilistic topic model for recipe representation. Other recipe attributes are also incorporated into the model to discover cuisine-course specific topic distribution to enable both cuisine and ingredient oriented  recipe recommendation. Lin \textit{et al.}~\cite{Lin-CBMF-KDDM2014} fused features from different types of information, such as ingredients, dietary, preparation, courses, cuisines and occasion into unified representation, and then used a content-driven matrix factorization approach to model the latent dimension of recipes, users, and features, respectively.

Combination of different  food modalities  leads to the multimodality of food, which particularly lies in the following two aspects: (1) One important factor determining our food choice is how we perceive food from its certain characteristics. We perceive food through a number of simultaneous sensory streams-we see food objects, hear its sounds when chewing, feel its texture, smell its odors and taste its flavors. Therefore, food perception actually involves multi-modalities.  Food perception has an important affect on food preference learning. Accurate food perception should also learn  multimodal sensory information \cite{Verhagen2006The}. (2) We are living in the age of the social web, and leave digital traces of all types of food-relevant activities online. For example, online recipe websites often have rich modalities and metadata about recipes. Each food item in Yummly consists of the visual food photo, textual content (e.g., name and ingredients) and attributes (e.g., cuisine and course). With the advent of Twitter, Facebook and Instagram, it is a common practice to upload food relevant text, share food photos.
The multimodal information contains rich, detailed information about what people eat, how much, when, where, and even with whom. Such multimodal food information can provide a valuable signal for diet profiling and food preference elicitation.

Multimodal fusion methods \cite{wqmin-DRA-mm2017,Jing-CMR-MM2017,Min-YAWYE-TMM2018} are needed to effectively combine different modalities. For example, Nag \textit{et al.}~\cite{Nag-HML-ICMR2017} combined  food images and  GPS context for Foodlog. They~\cite{Nag-CMHSE-MM18} then fused multiple user  source data streams along with the domain knowledge for cross-modal health state estimation. Recently, Markus \textit{et al.} \cite{Markus-Recipe-ICWSM2018} used different kinds of features from different modalities, including a recipe's title, ingredient list and cooking directions, popularity indicators (e.g., the number of ratings) and visual features to estimate the healthiness of recipes for recipe recommendation. In addition, Chu \textit{et al.}~\cite{Chu-HRS-WWW2017} combined text information, metadata and  visual features for restaurant attributes and user preference representation. Two common recommendation approaches, i.e., content-based filtering and collaborative filtering are then integrated for restaurant recommendation. Recently, the capability of Deep Learning (DL) in processing heterogeneous data ~\cite{WeiqingMin-BSC-TMM2017,Salvador-LCME-CVPR2017}  brings more opportunities in recommending more accurate and diverse items. For example, Min \textit{et al.}~\cite{WeiqingMin-BSC-TMM2017} utilized the deep belief network to jointly model visual information, textual content (e.g., ingredients), and attributes (e.g., cuisine and course) to solve recipe-oriented problems. Salvador \textit{et al.}~\cite{Salvador-LCME-CVPR2017} developed a multi-modal deep neural model which jointly learns to embed images and recipes in a common space for shared multi-modal representation.

\begin{figure}
\centering
\includegraphics[width=0.45\textwidth]{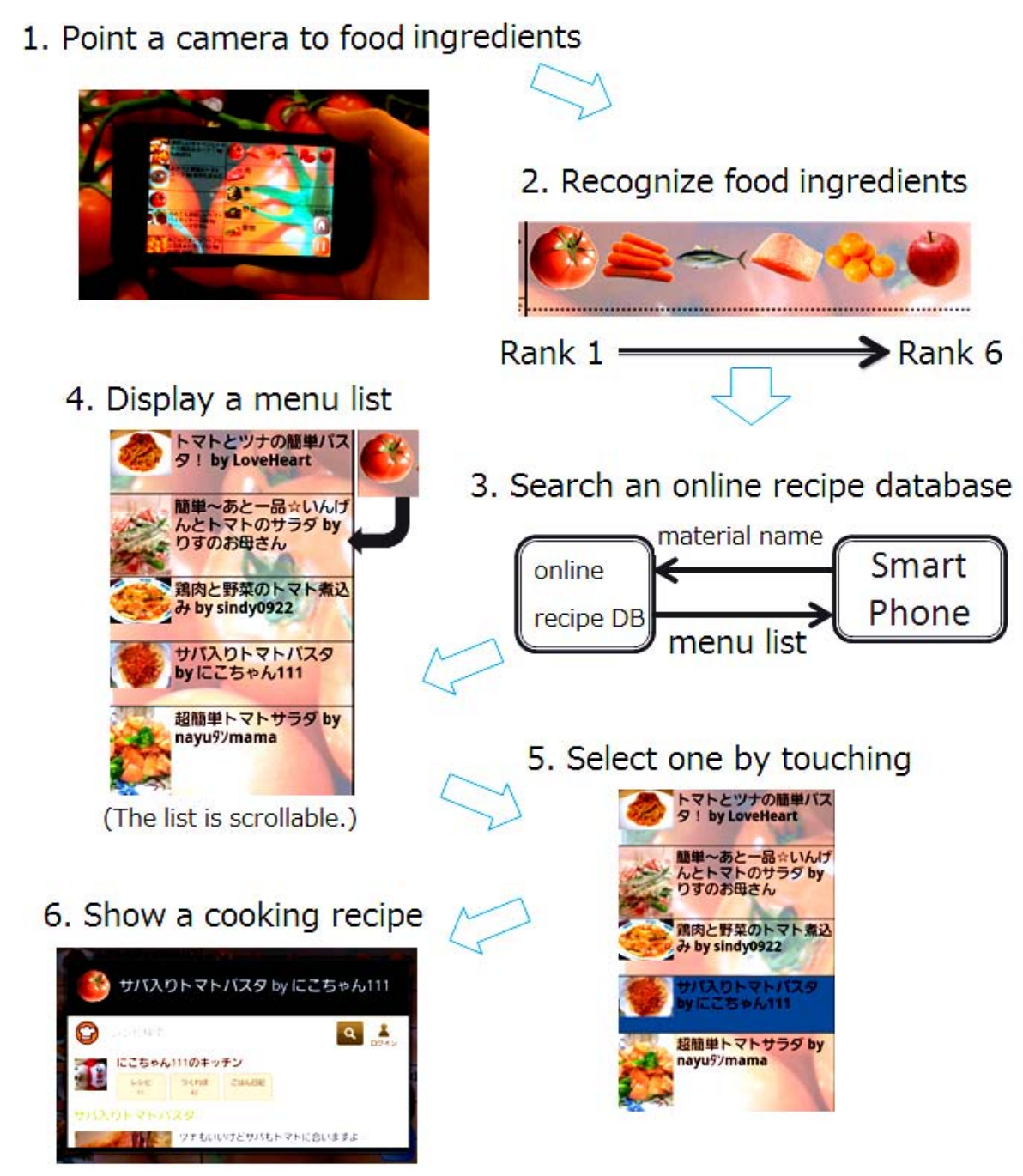}
\caption{Illustration of the overall recipe recommendation system~\cite{Kawano-mirurecipe-ICMEW2013}.}
\label{recipe-recommendation}
\end{figure}

As one representative work, Fig.~\ref{recipe-recommendation}  shows the flow of recipe recommendation system from the smart phone \cite{Kawano-mirurecipe-ICMEW2013,Maruyama-RMRR-MM2012}. The user first points a smartphone camera toward food ingredients at a grocery store or at a kitchen to recognize food ingredients, and then searches online cooking recipe databases with the name of the recognized food ingredient as a search keyword, and retrieve a menu list. Finally, the user selects one menu from the menu list and  displays the corresponding cooking recipe including a list on necessary ingredients and seasonings and a cooking procedure on the pop-up window.

Compared with multimodal data, KG can organize heterogeneous food data into one effective semantic structure to enable food recommendation. It has been widely used for news recommendation~\cite{Hongwei-DKN-WWW2018} and movie recommendation~\cite{Zhang-CKERS-SIGKDD2016}. To our knowledge, there is only one work~\cite{Zulaika-EPCARFS-Proceedings2018}, which constructed the food KG for semantic recipe retrieval.

Discussion. Rich food information is generally heterogeneous and complex, and is roughly divided into two types: multi-modal data and KG. Multimodal food analysis  is becoming one fundamental stage for food recommendation. Deep learning methods  have been one popular method  for multimodal food analysis. However, some traditional models, such as probabilistic graphical models  are still being used, especially for tasks with limited training samples. Despite these advances,  multimodal food analysis still faces a number of difficulties. For example, some modalities such as smell and taste are very important for food recommendation, but is harder to quantify, not to mention the following multimodal fusion. In addition, there are few KG-oriented food recommendation methods because of the lack of a large-scale food KG. Constructing the food KG to enable food recommendation will help improve the recommendation performance and interpretability.

\subsection{Nutrition and Health Oriented Food Recommendation}
Many people are facing the problem of making healthier food decisions to reduce the risk of chronic diseases such as obesity and diabetes, which are very relevant to what we eat. Therefore,  food recommendation  not only caters  user's food preference but should be also able to take  user's health into account. Correspondingly, how to build the model to balance these two components becomes the core problem for health and nutrient-based food recommendation.

Most methods have tried to incorporate healthiness into the recommendation process by substituting ingredients ~\cite{Teng-RRUIN-WSC2012,Harvey2017Exploiting}, incorporating calorie counts \cite{Ge-HFRS-Rec2015}, generating food plans \cite{Elsweiler-TAMPR-RecSys15}, and incorporating  nutritional facts~\cite{Nag-HML-ICMR2017,Ng-PRRT-MEDES2017,NagPSLWJ17}. For example, Ge  \textit{et al.}  \cite{Ge-HFRS-Rec2015} simply calculated the weighted mean between the preference component and health component, where the  weights are manually adjusted by the user. Nag \textit{et al.}~\cite{Nag-HML-ICMR2017} utilized food recognition method to obtain  food-relevant information, and then obtain nutritional information for recommendation. In addition to health-aware food recommendation  for general public, there are some works on  recommendation for toddlers. For example, Yiu-Kai \textit{et al.}~\cite{Ng-PRRT-MEDES2017} proposed a personalized toddler recipe recommendation system which incorporates the standard nutrition guideline published by the US government  with users' food preference for toddlers. A  hybrid recommendation approach, which incorporates content information directly as a linear constraint to provide additional insights about the contents themselves is adopted. Ribeiro  \textit{et al.}~\cite{David-SousChef-ICT4AWE2017} considers more factors including nutrition, food preferences and the budget for meal recommender system for older adults.

Discussion. One important difference between food recommendation and other types of recommendation is that users' health information should be considered. Current methods mainly  consider the nutrition information via visual food analysis or external nutrition table look-up for nutrition-oriented recommendation. However, user's real-time health-relevant state information is neglected. With the development of various sensor technologies, such information can be obtained from various portable devices, such as watches and bracelets. In addition, existing methods balance user's food preference and  user's health via simple fusion, such as weighted summation~\cite{Ge-HFRS-Rec2015}, which are probably not accurate. More effective non-linear fusion methods between these two factors should be explored.

\section{Research Challenges}\label{Challenge}
Food recommendation recently has received more attention  for its potential applications in human health. Thus, it is  important to discuss existing challenges that form  major obstacles to current progress. This section presents key unresolved issues.
\subsection{Various Sensor Signal Fusion}
Compared with  recommendation in other fields,  besides common spatio-temporal context, we should  understand and model people states in food recommendation based on  multiple signals, such as the affective state, physical state and health state captured from various sensors.
The context information  is with different types of distribution. Therefore, a joint model is necessary to model the user state  to accommodate a mixture of both discrete and continuous context variables.  In addition, although we can resort  sensors to obtain various context, there are still many signals, which are harder or inconvenient to obtain. The measure inaccuracy, uncertainty and unreliability from these sensors also exist. Therefore, further innovations on sensors are also needed for materials and devices, with the aim of more accurate sensing, the capability of detecting more context and  real-time monitoring.  In addition, how to  filter out noise  and conduct effective context fusion from these sensors is worth studying in the future. When the semantics from  different modalities are not consistent, the conflict disambiguation and reasonable decision mechanism are further needed, but ignored in existing solutions.
\subsection{Personal Model Construction}
Few works focus on  personal model construction for food recommendation. Simple person model can be  defined by user's context and some inherent health parameters such as weight, height and  activity steps. In order to achieve comprehensive and accurate personal model construction, according to the objective self \cite{Jain-IEEEMM2014}, we should complete it through data ingestion, life event recognition, and pattern recognition. However, each component is hard to achieve and the  challenges derive from three-fold: (1) data ingestion uses various sensors (e.g., smartphones and various wearable devices) and preprocessing modules to extract appropriate attributes from raw sensor measurements. However, the information obtained from different sensors varies in many aspects. Methods to convert data to information and the reliability of information could be entirely different for different sensors. Furthermore, as new sensing technologies emerge and are now becoming omnipresent in daily lives,  more types of sensor data will be generated. It is difficult to effecitvely fuse heterogeneous multimedia information from different sensors via a unified model. (2) the proliferation of recipe-sharing websites (e.g., Yummly, Meishijie, foodspotting and Allrecipes) has resulted in huge user-uploaded food data collections. These recipes are associated with rich modality and attribute information. Such recipe data with rich types can be exploited to answer various food related questions in food recommendation. Besides recipe-sharing websites, the social media, such as Twitter, Foursquare, Flickr, and Instagram also provide large-scale food data uploaded by users.
An increasing amount of user-shared food-related data presents researchers with more opportunities for personal model construction. However, it also presents researchers with  challenges, such as much noise and the sheer size of user-shared food data. (3) accurate personal model construction benefits from  fast development of many technologies, such as natural-language processing, machine learning and computer vision  methods. For example, activity recognition techniques are necessary for life event recognition, and is still one hot problem in the computer vision. Lower recognition performance leads to inaccurate personal model construction, which conversely affects the experience of food recommendation.
\subsection{Visual Food Analysis}
Visual food analysis can obtain a high-level understanding of the type (e.g.,the food category and ingredients), the amount of food consumed by the user and even the calorie,  and thus  is very essential for  food recommendation. This category can broadly be divided into different types, such as food category recognition~\cite{Bossard-Food101-ECCV2014,Min-IG-CMAN-MM2019,Jiang-MSMVDFA-TIP2019}, food ingredient recognition~\cite{Maruyama-RMRR-MM2012}, cooking instruction recognition~\cite{Salvador-LCME-CVPR2017} and food quantity estimation~\cite{Meyers-Im2Calories-ICCV2015,Dehais2017Two}. However, accurate visual food analysis is very challenging. Food images have their own distinctive properties. A large number of food dishes have deformable food appearance and thus lacks rigid structures. Therefore, it is hard for us to achieve satisfactory analysis results based on existing visual analysis methods. In addition, visual food analysis involves fine-grained visual feature representation learning. However, we can not simply  use existing fine-grained feature learning methods for food analysis. One important reason is that existing fine-grained feature learning methods  generally assume that there are fixed semantic patterns in each image and the task is to discover these patterns. However, the concepts of common semantic patterns do not exist in many food images.  Therefore, we should design a new visual feature learning paradigm particularly for food.
\subsection{Multi-Modality Food Analysis}
Multi-modality is commonly the case in food recommendation. For example, each recipe item in Yummly consists of the food image, textual content (e.g.,ingredients and cooking instructions), attributes (e.g., cuisine and taste), calories and nutrition facts. Users adopt both captured food images and text information for diet recording in the Foodlog. How to effectively integrate various types of  information into the recommendation algorithm according to the characteristics of the specific recommendation scene has been a  challenging problem, and how to enable  effective feature extraction and fusion from various types of multimedia content information is also one core problem in this research field, not to mention the tight coupling between multi-modal fusion and collaborative filtering.

\section{Future Directions}\label{Directions}
Food recommendation is still in its infancy in the multimedia community, and much more needs to be done to promote the development in this field. In this section, we discuss some open research directions and new research perspectives of food recommendation.

\subsection{Food Preference Learning}
Compared with user's preference on other objects, user's food preference has its own special features. Factors that guide food preference are various, and it is a result of  a combination of  biological, psychological, social, cultural and historical influences. Benefiting from the rapidly growing body of publicly available food data, it is convenient to utilize these data to learn user's  food preference. However, they mainly  focused on modeling one or several aspects from these data and neglected many other factors, which are probably hard to obtain or model. Borrowing methodologies from other areas such as psychology and neuroscience probably provides new perspectives for accurate food preference learning. In addition, most existing food recommender systems  assumed that  user's food preference is  static. However,  users' food preferences are dynamic  in reality \cite{DFP-FQP2004}. A reasonable solution is to adopt  deep reinforcement learning to model temporal dynamics of food preferences for food recommendation.

\subsection{Large-Scale Benchmark Dataset Construction}
Like MovieLens-1M (\url{https://grouplens.org/datasets/movielens/1m/}) for movie recommendation, a large-scale dataset with rich user-food interaction and multimodal content information (e.g., food images, ingredients and user comments for one dish)  is also a critical resource for developing advanced  food recommendation  algorithms, as well as for providing critical training and benchmark data for such algorithms. Although some datasets for food recommendation \cite{Lin-CBMFM-KDDM2014} are available, there are few public large-scale datasets, especially with multimodal information (e.g., food images and other rich attributes) for food recommendation. Therefore, collecting and releasing a large-scale multimodal benchmark dataset for  the food recommendation task is in urgent need.

\subsection{Large-scale Food KG Construction}
A complete KG can provide a deeper and longer range of associations between items. It has been proved effectively in recommendation system for many fields, such as movies and books. Unfortunately, there is no food-oriented KG available. Therefore, we should construct food-oriented KG. In order to obtain comprehensive food profile, we should crawl and parse metadata  from various media sites with structured and semi-structured data. In many websites, metadata associated with content is poorly structured without a well defined ontology, especially in various social media. As a result, it is very hard to explore. Towards the food KG, we should define nodes and edges, and establish the standard for the food domain. A good starting point is to utilize food items from Wikipedia\footnote{\url{https://en.wikipedia.org/wiki/Lists_of_foods}} to build the food KG. In addition, we can even  associate visual information  and other modality information with this KG to build multimodal food knowledge graph.

\subsection{Tight Combination between DL and KG}
Like other types of recommendation, food recommendation systems work on a lot of heterogeneous multi-modal information. Current systems mostly leverage heterogeneous information  to improve the recommendation performance, while a lot of research efforts are needed regarding how to jointly leverage heterogeneous information  for food recommendation. These include a wide range of research tasks such as multi-modal  alignment on multiple different modalities and transfer learning over heterogeneous information sources for food recommendation. Recently, recommendation has undergone a fundamental paradigm shift towards using deep learning as a general-purpose solution. One key advantage of adopting DL  is that  arbitrary continuous and categorical features can be easily added to the model. some  works \cite{Yang2017Yum} have utilized   deep learning  for food recommendation. We expect the wide use of deep learning to enhance the performance of food recommendation.

On the other hand, incorporating KG into recommender systems has attracted increasing attention in recent years. Compared with multimodal information, by exploring the relationships within a KG, the connectivity between users and items can be discovered and provides rich and complementary information to user-item interactions.  However, to our knowledge, there is no work on food recommendation. The probable reason is that there is no constructed food-specific KG. Constructing the KG in the food domain requires us carefully define  nodes and  types of relationships, which is not easy to complete. Even we have available food KG, how to effectively combine KG embedding and collaborative filtering in the unified framework is still one hot topic and needs further exploration. Finally, effectively combining DL and KG jointly for food recommendation in a two-wheeled driven way will be needed.
\subsection{Explainable Food Recommendation}
Explainable recommendation refers to personalized  recommendation algorithms that address the problem of why-they not only provide users with the recommendations, but also provide explanations to make
the user  aware of why such items are recommended.  The same is true for food recommendation. Because food recommendation considers more complex and various factors, in this way, explainable food recommendation helps to find out main factors affecting recommended results and thus improve  user satisfaction of recommendation systems. Incorporating KG into food recommendation is a promising direction for explainable food recommendation. In addition, the research community has  been leveraging deep learning techniques for food recommendation. Current approaches focus on designing deep models to improve the performance. However, the research of leveraging deep models for explainable food recommendation is still in its initial stage, and there is much more to be explored in the future.

\section{Conclusions}
\label{Conclusions}
Food recommendation is a promising and important research direction
for its importance to quality of life for people and potential
applications in human health. It
provides people with ranked food items using rich context and
knowledge, personal model constructed dynamically using Foodlog, and
heterogeneous food analysis to understand nutrition and taste characteristics.
However, there are very few
systematic reviews.  We believe that a critical and exhaustive review presented here
may encourage researchers to shape this area.  We provide
a comprehensive and in-depth summary of current efforts as well as
detail open problems in this area. This paper contains a
comprehensive overview of food recommendation by proposing and using a unified
food recommendation framework. Main issues affecting food recommendation are
identified. The existing solutions are also introduced.

The area of food recommendation is still in its infancy with
many challenges and open questions. Many of the challenges
cannot be addressed using techniques from only one discipline.
We believe that multidisciplinary research that combines
nutrition, food science, psychology, biology, anthropology and
sociology will lead to more powerful approaches
and technologies to handle food recommendation. Thus, considering
huge potentials in human health and great commercial
applications, we will look forward to seeing the surge in this
research field in the foreseeable future.



\bibliographystyle{IEEEtran}
\bibliography{FR_TMM19}  

\begin{thebibliography}{10}
\providecommand{\url}[1]{#1}
\csname url@samestyle\endcsname
\providecommand{\newblock}{\relax}
\providecommand{\bibinfo}[2]{#2}
\providecommand{\BIBentrySTDinterwordspacing}{\spaceskip=0pt\relax}
\providecommand{\BIBentryALTinterwordstretchfactor}{4}
\providecommand{\BIBentryALTinterwordspacing}{\spaceskip=\fontdimen2\font plus
\BIBentryALTinterwordstretchfactor\fontdimen3\font minus
  \fontdimen4\font\relax}
\providecommand{\BIBforeignlanguage}[2]{{%
\expandafter\ifx\csname l@#1\endcsname\relax
\typeout{** WARNING: IEEEtran.bst: No hyphenation pattern has been}%
\typeout{** loaded for the language `#1'. Using the pattern for}%
\typeout{** the default language instead.}%
\else
\language=\csname l@#1\endcsname
\fi
#2}}
\providecommand{\BIBdecl}{\relax}
\BIBdecl

\bibitem{Tirosheaav0120}
A.~Tirosh, E.~S. Calay, G.~Tuncman, K.~C. Claiborn, K.~E. Inouye, K.~Eguchi,
  M.~Alcala, M.~Rathaus, K.~S. Hollander, I.~Ron, R.~Livne, Y.~Heianza, L.~Qi,
  I.~Shai, R.~Garg, and G.~S. Hotamisligil, ``The short-chain fatty acid
  propionate increases glucagon and {FABP4} production, impairing insulin
  action in mice and humans,'' \emph{Science Translational Medicine}, vol.~11,
  no. 489, 2019.

\bibitem{Federation-IDF2015}
I.D.Federation, \emph{IDF Diabetes Atlas}.\hskip 1em plus 0.5em minus
  0.4em\relax International Diabetes Federation, 2015.

\bibitem{GBD-Global-Lancet2018}
G.~. R.~F. Collaborators, ``Global, regional, and national incidence,
  prevalence, and years lived with disability for 354 diseases and injuries for
  195 countries and territories, 1990-2017: a systematic analysis for the
  global burden of disease study 2017,'' \emph{The Lancet}, vol. 392, no.
  10159, pp. 1789 -- 1858, 2018.

\bibitem{Min2018A}
W.~Min, S.~Jiang, L.~Liu, Y.~Rui, and R.~Jain, ``A survey on food computing,''
  \emph{ACM Comput. Surv.}, vol.~52, no.~5, pp. 92:1--92:36, 2019.

\bibitem{Weiqing-CAMVR-MSJ2017}
W.~Min, S.~Jiang, S.~Wang, R.~Xu, Y.~Cao, L.~Herranz, and Z.~He, ``A survey on
  context-aware mobile visual recognition,'' \emph{Multimedia Systems},
  vol.~23, no.~6, pp. 647--665, 2017.

\bibitem{Boll-HM-IEEEMM2018}
S.~Boll, J.~Meyer, and N.~E. O'Connor, ``Health media: From multimedia signals
  to personal health insights,'' \emph{IEEE MultiMedia}, vol.~25, no.~1, pp.
  51--60, 2018.

\bibitem{Maruyama-RMRR-MM2012}
T.~Maruyama, Y.~Kawano, and K.~Yanai, ``Real-time mobile recipe recommendation
  system using food ingredient recognition,'' in \emph{Proceedings of the ACM
  international workshop on interactive multimedia on mobile and portable
  devices}, 2012, pp. 27--34.

\bibitem{Nag2017Live}
N.~Nag, V.~Pandey, and R.~Jain, ``Live personalized nutrition recommendation
  engine,'' in \emph{Proceedings of the 2Nd International Workshop on
  Multimedia for Personal Health and Health Care}, 2017, pp. 61--68.

\bibitem{Nag-HML-ICMR2017}
------, ``Health multimedia: Lifestyle recommendations based on diverse
  observations,'' in \emph{Proceedings of the ACM on International Conference
  on Multimedia Retrieval}, 2017, pp. 99--106.

\bibitem{Trattner2017Food}
C.~Trattner and D.~Elsweiler, ``Food recommender systems: Important
  contributions, challenges and future research directions,'' \emph{arXiv
  preprint arXiv:1711.02760}, 2017.

\bibitem{Burke2002Hybrid}
R.~Burke, ``Hybrid recommender systems: Survey and experiments,'' \emph{User
  Modeling and User-Adapted Interaction}, vol.~12, no.~4, pp. 331--370, 2002.

\bibitem{Lou2017Recent}
Z.~Lou, Li, L.~Wang, and G.~Shen, ``Recent progress of self-powered sensing
  systems for wearable electronics,'' \emph{Small}, vol.~13, no.~45, p.
  1701791, 2017.

\bibitem{Strickland-Sensors-IEEESp2018}
E.~Strickland, ``3 sensors to track every bite and gulp [news],'' \emph{IEEE
  Spectrum}, vol.~55, no.~7, pp. 9--10, 2018.

\bibitem{Aizawa-FoodLog-IEEEMM2015}
K.~Aizawa and M.~Ogawa, ``Foodlog: Multimedia tool for healthcare
  applications,'' \emph{IEEE MultiMedia}, vol.~22, no.~2, pp. 4--8, 2015.

\bibitem{Kim2018Nutritionally}
S.~Kim, M.~F. Fenech, and P.~J. Kim, ``Nutritionally recommended food for semi-
  to strict vegetarian diets based on large-scale nutrient composition data,''
  \emph{Scientific Reports}, vol.~8, no.~1, p. 4344, 2018.

\bibitem{Elahi2017User}
M.~Elahi, D.~Elsweiler, G.~Groh, M.~Harvey, B.~Ludwig, F.~Ricci, and A.~Said,
  ``User nutrition modelling and recommendation: Balancing simplicity and
  complexity,'' in \emph{Adjunct Publication of the Conference on User
  Modeling, Adaptation and Personalization}, 2017, pp. 93--96.

\bibitem{Yang2017Yum}
L.~Yang, C.~K. Hsieh, H.~Yang, J.~P. Pollak, N.~Dell, S.~Belongie, C.~Cole, and
  D.~Estrin, ``{Yum-Me}: A personalized nutrient-based meal recommender
  system,'' \emph{ACM Transactions on Information Systems}, vol.~36, no.~1,
  p.~7, 2017.

\bibitem{Bossard-Food101-ECCV2014}
L.~Bossard, M.~Guillaumin, and L.~Van~Gool, ``Food-101-mining discriminative
  components with random forests,'' in \emph{European Conference on Computer
  Vision}, 2014, pp. 446--461.

\bibitem{Smith2014The}
L.~Smith and M.~Gasser, ``The development of embodied cognition: six lessons
  from babies.'' \emph{Artificial Life}, vol.~11, no. 1-2, pp. 13--29, 2014.

\bibitem{Cordeiro-RMFJ-HFCS2015}
F.~Cordeiro, E.~Bales, E.~Cherry, and J.~Fogarty, ``Rethinking the mobile food
  journal: Exploring opportunities for lightweight photo-based capture,'' in
  \emph{Proceedings of the ACM Conference on Human Factors in Computing
  Systems}, 2015, pp. 3207--3216.

\bibitem{Yanjie-UPL-SIAM2014}
Y.~Fu, B.~Liu, G.~Yong, Z.~Yao, and X.~Hui, ``User preference learning with
  multiple information fusion for restaurant recommendation,'' in
  \emph{Proceedings of the SIAM International Conference on Data Mining}, 2014,
  p. 470–478.

\bibitem{NagPSLWJ17}
N.~Nitish, P.~Vaibhav, S.~Abhisaar, L.~Jonathan, W.~Runyi, and J.~Ramesh,
  ``Pocket dietitian: automated healthy dish recommendations by location,'' in
  \emph{International Conference on Image Analysis and Processing}, 2017, pp.
  444--452.

\bibitem{Sanjo2017Recipe}
S.~Sanjo and M.~Katsurai, ``Recipe popularity prediction with deep
  visual-semantic fusion,'' in \emph{Proceedings of the ACM on Conference on
  Information and Knowledge Management}, 2017, pp. 2279--2282.

\bibitem{Meyer-DHDE-DPC2014}
J.~{Meyer} and S.~{Boll}, ``Digital health devices for everyone!'' \emph{IEEE
  Pervasive Computing}, vol.~13, no.~2, pp. 10--13, 2014.

\bibitem{Fulgoni-MNQF-JN2009}
I.~Fulgoni, Victor~L., D.~R. Keast, and A.~Drewnowski, ``{Development and
  Validation of the Nutrient-Rich Foods Index: A Tool to Measure Nutritional
  Quality of Foods},'' \emph{The Journal of Nutrition}, vol. 139, no.~8, pp.
  1549--1554, 2009.

\bibitem{Julia-VFSA-EJN2016}
C.~Julia, C.~M{\'e}jean, M.~Touvier, S.~P{\'e}neau, C.~Lassale, P.~Ducrot,
  S.~Hercberg, and E.~Kesse-Guyot, ``Validation of the fsa nutrient profiling
  system dietary index in french adults---findings from suvimax study,''
  \emph{European Journal of Nutrition}, vol.~55, no.~5, pp. 1901--1910, 2016.

\bibitem{Jain-IEEEMM2014}
R.~{Jain} and L.~{Jalali}, ``Objective self,'' \emph{IEEE MultiMedia}, vol.~21,
  no.~4, pp. 100--110, 2014.

\bibitem{Freyne-IFP-ICIUI2010}
J.~Freyne and S.~Berkovsky, ``Intelligent food planning: Personalized recipe
  recommendation,'' in \emph{Proceedings of the International Conference on
  Intelligent User Interfaces}, 2010, pp. 321--324.

\bibitem{Ge2015Using}
M.~Ge, M.~Elahi, I.~Ferna\'{a}ndez-Tob\'{\i}as, F.~Ricci, and D.~Massimo,
  ``Using tags and latent factors in a food recommender system,'' in
  \emph{Proceedings of the 5th International Conference on Digital Health},
  2015, pp. 105--112.

\bibitem{Yang-PlateClick-MM2015}
L.~Yang, Y.~Cui, F.~Zhang, J.~P. Pollak, S.~Belongie, and D.~Estrin,
  ``Plateclick: Bootstrapping food preferences through an adaptive visual
  interface,'' in \emph{Proceedings of the 24th ACM International on Conference
  on Information and Knowledge Management}, 2015, pp. 183--192.

\bibitem{Gao-HANVFR-TMM2019}
X.~{Gao}, F.~{Feng}, X.~{He}, H.~{Huang}, X.~{Guan}, C.~{Feng}, Z.~{Ming}, and
  T.~{Chua}, ``Hierarchical attention network for visually-aware food
  recommendation,'' \emph{IEEE Transactions on Multimedia}, pp. 1--1, 2019.

\bibitem{Zeng-Zeng-MHFI18}
J.~Zeng, Y.~I. Nakano, T.~Morita, I.~Kobayashi, and T.~Yamaguchi, ``Eliciting
  user food preferences in terms of taste and texture in spoken dialogue
  systems,'' in \emph{Proceedings of the 3rd International Workshop on
  Multisensory Approaches to Human-Food Interaction}, 2018, pp. 6:1--6:5.

\bibitem{Hyungik-MFJ-ACMMM2018}
H.~Oh, J.~Nguyen, S.~Soundararajan, and R.~Jain, ``Multimodal food
  journaling,'' in \emph{Proceedings of the 3rd International Workshop on
  Multimedia for Personal Health and Health Care, HealthMedia@MM 2018}, 2018,
  pp. 39--47.

\bibitem{Meyers-Im2Calories-ICCV2015}
A.~Meyers, N.~Johnston, V.~Rathod, A.~Korattikara, A.~Gorban, N.~Silberman,
  S.~Guadarrama, G.~Papandreou, J.~Huang, and K.~P. Murphy, ``{Im2Calories}:
  Towards an automated mobile vision food diary,'' in \emph{Proceedings of the
  IEEE International Conference on Computer Vision}, 2015, pp. 1233--1241.

\bibitem{Wang2016Where}
H.~Wang, W.~Min, X.~Li, and S.~Jiang, ``Where and what to eat: Simultaneous
  restaurant and dish recognition from food image,'' in \emph{Pacific Rim
  Conference on Multimedia}, 2016, pp. 520--528.

\bibitem{Horiguchi-PCFIR-TMM2018}
S.~{Horiguchi}, S.~{Amano}, M.~{Ogawa}, and K.~{Aizawa}, ``Personalized
  classifier for food image recognition,'' \emph{IEEE Transactions on
  Multimedia}, vol.~20, no.~10, pp. 2836--2848, 2018.

\bibitem{Aizawa-FBE-TMM2013}
K.~Aizawa, Y.~Maruyama, H.~Li, and C.~Morikawa, ``Food balance estimation by
  using personal dietary tendencies in a multimedia food log,'' \emph{IEEE
  Transactions on Multimedia}, vol.~15, no.~8, pp. 2176--2185, 2013.

\bibitem{XuRuihan-GMDR-TMM2015}
R.~Xu, L.~Herranz, S.~Jiang, S.~Wang, X.~Song, and R.~Jain, ``Geolocalized
  modeling for dish recognition,'' \emph{IEEE Transactions on Multimedia},
  vol.~17, no.~8, pp. 1187--1199, 2015.

\bibitem{Chen-DIRCRR-MM2016}
J.~Chen and C.-W. Ngo, ``Deep-based ingredient recognition for cooking recipe
  retrieval,'' in \emph{Proceedings of the ACM on Multimedia Conference}, 2016,
  pp. 32--41.

\bibitem{Harvey2017Exploiting}
M.~Harvey, M.~Harvey, and M.~Harvey, ``Exploiting food choice biases for
  healthier recipe recommendation,'' in \emph{International ACM SIGIR
  Conference on Research and Development in Information Retrieval}, 2017, pp.
  575--584.

\bibitem{Ming-FPR-MMM2018}
Z.-Y. Ming, J.~Chen, Y.~Cao, C.~Forde, C.-W. Ngo, and T.~S. Chua, ``Food photo
  recognition for dietary tracking: System and experiment,'' in
  \emph{MultiMedia Modeling}, 2018, pp. 129--141.

\bibitem{Dehais2017Two}
J.~Dehais, M.~Anthimopoulos, S.~Shevchik, and S.~Mougiakakou, ``Two-view {3D}
  reconstruction for food volume estimation,'' \emph{IEEE Transactions on
  Multimedia}, vol.~19, no.~5, pp. 1090--1099, 2017.

\bibitem{Forbes-CBMF-RS2011}
P.~Forbes and M.~Zhu, ``Content-boosted matrix factorization for recommender
  systems: experiments with recipe recommendation,'' in \emph{Proceedings of
  the fifth ACM conference on Recommender systems}, 2011, pp. 261--264.

\bibitem{Devis-SDFR-WISE2015}
D.~Bianchini, V.~De~Antonellis, and M.~Melchiori, ``A web-based application for
  semantic-driven food recommendation with reference prescriptions,'' in
  \emph{Web Information Systems Engineering}.\hskip 1em plus 0.5em minus
  0.4em\relax Cham: Springer International Publishing, 2015, pp. 32--46.

\bibitem{Min-YAWYE-TMM2018}
W.~Min, B.-K. Bao, S.~Mei, Y.~Zhu, Y.~Rui, and S.~Jiang, ``You are what you
  eat: Exploring rich recipe information for cross-region food analysis,''
  \emph{IEEE Transactions on Multimedia}, vol.~20, no.~4, pp. 950--964, 2018.

\bibitem{Lin-CBMF-KDDM2014}
C.-J. Lin, T.-T. Kuo, and S.-D. Lin, ``A content-based matrix factorization
  model for recipe recommendation,'' in \emph{Advances in Knowledge Discovery
  and Data Mining}.\hskip 1em plus 0.5em minus 0.4em\relax Cham: Springer
  International Publishing, 2014, pp. 560--571.

\bibitem{Verhagen2006The}
J.~V. Verhagen and L.~Engelen, ``The neurocognitive bases of human multimodal
  food perception: sensory integration.'' \emph{Neuroscience \& Biobehavioral
  Reviews}, vol.~30, no.~5, pp. 613--50, 2006.

\bibitem{wqmin-DRA-mm2017}
W.~Min, S.~Jiang, S.~Wang, J.~Sang, and S.~Mei, ``A delicious recipe analysis
  framework for exploring multi-modal recipes with various attributes,'' in
  \emph{Proceedings of the 2017 {ACM} on Multimedia Conference}, 2017, pp.
  402--410.

\bibitem{Jing-CMR-MM2017}
J.~Chen, C.~Ngo, and T.~Chua, ``Cross-modal recipe retrieval with rich food
  attributes,'' in \emph{Proceedings of the ACM on Multimedia Conference},
  2017, pp. 1771--1779.

\bibitem{Nag-CMHSE-MM18}
N.~Nag, V.~Pandey, P.~J. Putzel, H.~Bhimaraju, S.~Krishnan, and R.~Jain,
  ``Cross-modal health state estimation,'' in \emph{Proceedings of the 26th ACM
  International Conference on Multimedia}, 2018, pp. 1993--2002.

\bibitem{Markus-Recipe-ICWSM2018}
M.~Rokicki, C.~Trattner, and E.~Herder, ``The impact of recipe features, social
  cues and demographics on estimating the healthiness of online recipes,'' in
  \emph{Proceedings of the Twelfth International Conference on Web and Social
  Media}, 2018, pp. 310--319.

\bibitem{Chu-HRS-WWW2017}
W.~T. Chu and Y.~L. Tsai, ``A hybrid recommendation system considering visual
  information for predicting favorite restaurants,'' \emph{World Wide
  Web-internet \& Web Information Systems}, vol.~20, no.~6, pp. 1313--1331,
  2017.

\bibitem{WeiqingMin-BSC-TMM2017}
W.~Min, S.~Jiang, J.~Sang, H.~Wang, X.~Liu, and L.~Herranz, ``Being a
  supercook: Joint food attributes and multi-modal content modeling for recipe
  retrieval and exploration,'' \emph{IEEE Transactions on Multimedia}, vol.~19,
  no.~5, pp. 1100 -- 1113, 2017.

\bibitem{Salvador-LCME-CVPR2017}
A.~Salvador, N.~Hynes, Y.~Aytar, J.~Marin, F.~Ofli, I.~Weber, and A.~Torralba,
  ``Learning cross-modal embeddings for cooking recipes and food images,'' in
  \emph{Computer Vision and Pattern Recognition}, 2017, pp. 3020--3028.

\bibitem{Kawano-mirurecipe-ICMEW2013}
Y.~{Kawano}, T.~{Sato}, T.~{Maruyama}, and K.~{Yanai}, ``mirurecipe: A mobile
  cooking recipe recommendation system with food ingredient recognition,'' in
  \emph{IEEE International Conference on Multimedia and Expo Workshops}, 2013,
  pp. 1--2.

\bibitem{Hongwei-DKN-WWW2018}
H.~Wang, F.~Zhang, X.~Xie, and M.~Guo, ``{DKN:} deep knowledge-aware network
  for news recommendation,'' in \emph{Proceedings of the 2018 World Wide Web
  Conference on World Wide Web}, 2018, pp. 1835--1844.

\bibitem{Zhang-CKERS-SIGKDD2016}
F.~Zhang, N.~J. Yuan, D.~Lian, X.~Xie, and W.~Ma, ``Collaborative knowledge
  base embedding for recommender systems,'' in \emph{Proceedings of the 22nd
  {ACM} {SIGKDD} International Conference on Knowledge Discovery and Data
  Mining}, 2016, pp. 353--362.

\bibitem{Zulaika-EPCARFS-Proceedings2018}
U.~Zulaika, A.~Gutierrez, and D.~Lopez-de Ipina, ``Enhancing profile and
  context aware relevant food search through knowledge graphs,''
  \emph{Proceedings}, 2018.

\bibitem{Teng-RRUIN-WSC2012}
C.-Y. Teng, Y.-R. Lin, and L.~A. Adamic, ``Recipe recommendation using
  ingredient networks,'' in \emph{Proceedings of the ACM Web Science
  Conference}, 2012, pp. 298--307.

\bibitem{Ge-HFRS-Rec2015}
M.~Ge, F.~Ricci, and D.~Massimo, ``Health-aware food recommender system,'' in
  \emph{Proceedings of the Conference on Recommender Systems}, 2015, pp.
  333--334.

\bibitem{Elsweiler-TAMPR-RecSys15}
D.~Elsweiler and M.~Harvey, ``Towards automatic meal plan recommendations for
  balanced nutrition,'' in \emph{Proceedings of the 9th ACM Conference on
  Recommender Systems}, 2015, pp. 313--316.

\bibitem{Ng-PRRT-MEDES2017}
Y.-K. Ng and M.~Jin, ``Personalized recipe recommendations for toddlers based
  on nutrient intake and food preferences,'' in \emph{Proceedings of the 9th
  International Conference on Management of Digital EcoSystems}, 2017, pp.
  243--250.

\bibitem{David-SousChef-ICT4AWE2017}
D.~Ribeiro, J.~Ribeiro, M.~J.~M. Vasconcelos, E.~F. Vieira, and A.~C.
  de~Barros, ``Souschef: Improved meal recommender system for portuguese older
  adults,'' in \emph{Information and Communication Technologies for Ageing Well
  and e-Health}.\hskip 1em plus 0.5em minus 0.4em\relax Springer International
  Publishing, 2018, pp. 107--126.

\bibitem{Min-IG-CMAN-MM2019}
W.~Min, L.~Liu, Z.~Luo, and S.~Jiang, ``Ingredient-guided cascaded
  multi-attention network for food recognition,'' in \emph{Proceedings of the
  27th ACM International Conference on Multimedia}, 2019, pp. 1331--1339.

\bibitem{Jiang-MSMVDFA-TIP2019}
S.~{Jiang}, W.~{Min}, L.~{Liu}, and Z.~{Luo}, ``Multi-scale multi-view deep
  feature aggregation for food recognition,'' \emph{IEEE Transactions on Image
  Processing}, vol.~29, no.~1, pp. 265--276, 2019.

\bibitem{DFP-FQP2004}
J.~Delarue and E.~Loescher, ``Dynamics of food preferences: a case study with
  chewing gums,'' \emph{Food Quality and Preference}, vol.~15, no.~7, pp. 771
  -- 779, 2004.

\bibitem{Lin-CBMFM-KDDM2014}
C.-J. Lin, T.-T. Kuo, and S.-D. Lin, ``A content-based matrix factorization
  model for recipe recommendation,'' in \emph{Advances in Knowledge Discovery
  and Data Mining}.\hskip 1em plus 0.5em minus 0.4em\relax Springer
  International Publishing, 2014, pp. 560--571.

\end{thebibliography}

\end{document}